\documentclass[journal,draftcls,onecolumn,12pt,twoside]{IEEEtranTCOM}
\usepackage{graphicx}
\usepackage{amsmath,amsfonts,amssymb}
\usepackage{algorithmicx}
\usepackage{algorithm}
\usepackage{algpseudocode}
\usepackage{epstopdf}
\usepackage{subfigure}

\algnewcommand{\IIf}[1]{\State\algorithmicif\ #1\ \algorithmicthen}
\algnewcommand{\EndIIf}{\unskip\ \algorithmicend\ \algorithmicif}

\newtheorem{definition}{Definition}
\newtheorem{proposition}{Proposition}

\usepackage{xcolor}

\begin{document}
%
\author{\IEEEauthorblockN{L. Chandesris\IEEEauthorrefmark{2}\IEEEauthorrefmark{3},
V. Savin\IEEEauthorrefmark{2}, D. Declercq\IEEEauthorrefmark{3}  \\ {ludovic.chandesris@cea.fr,  valentin.savin@cea.fr,  declercq@ensea.fr}}\\
\IEEEauthorrefmark{2}CEA-LETI / Minatec, Grenoble, France \\
\IEEEauthorrefmark{3}ETIS, ENSEA/UCP/CNRS, Cergy-Pontoise, France}

\title{Dynamic-SCFlip Decoding of Polar Codes}

\markboth{}{}

\maketitle

\begin{abstract}
This paper proposes a generalization of the recently introduced Successive Cancellation Flip (SCFlip) decoding of polar codes, characterized by a number of extra decoding attempts, where one or several positions are flipped from the standard Successive Cancellation (SC) decoding. To make such an approach effective, we first introduce the concept of higher-order bit-flips, and propose a new metric to determine the bit-flips that are more likely to correct the trajectory of the SC decoding. We then propose a generalized SCFlip decoding algorithm, referred to as Dynamic-SCFlip (D-SCFlip), which  dynamically builds a list of candidate bit-flips, while guaranteeing that extra decoding attempts are performed by decreasing probability of success. Simulation results show that D-SCFlip is an effective alternative  to SC-List decoding of polar codes, by providing very good error correcting performance, with an average computation complexity close to the one of the SC decoder.
\end{abstract}


\section{Introduction}\label{sec:intro}

Polar codes are a recently discovered family of error correcting codes \cite{arikan2009channel}, known to achieve the capacity of any binary-input memoryless output-symmetric channel. Their construction relies on a specific recursive encoding procedure that synthesizes a set of $N$ virtual channels from $N$ instances of the transmission channel, where $N$ denotes the code-length. The recursive encoding procedure is reversed at the receiver end, by applying a Successive Cancellation (SC) decoder. The asymptotic effectiveness of the SC decoder derives from the fact that the synthesized channels tend to become either noiseless or completely noisy, as the code-length goes to infinity, phenomenon which is known as ``channel polarization''. However, for short to moderate code-lengths the incomplete polarization of the virtual channels may drastically penalize the error correction performance of the SC decoder. The main approaches proposed in the literature to address this issue rely on either modified kernels for the recursive encoding procedure, aimed at increasing the rate of polarization \cite{presman2011polar,miloslavskaya2012design}, or enhanced versions of the SC decoder \cite{tal2011list, afisiadis2014low, niu2012stack}, aimed at increasing its ability to deal with incompletely polarized channels. 

The SC-List (SCL) decoder proposed in \cite{tal2011list} significantly improves the error correction performance for short to moderate block lengths, and is also known to approach the Maximum-Likelihood (ML) decoding performance at high Signal to Noise Ratio (SNR). Moreover, to advantageously exploit the potential of SCL decoding, especially when the size of the decoded list is large, the concatenation of an outer Cyclic Redundancy Check (CRC) code has also been proposed in \cite{tal2011list}, to help identifying the correct message within the decoded list. Concatenated CRC-Polar codes under under SCL decoding is the best polar-coding system proposed so far, and has been shown to compete with other families of modern error correcting codes, such as Low Density Parity Check (LDPC) and Turbo codes. However, SCL decoder suffers from high storage and computational complexity, which grows linearly with the size of the list. Several improvements have been proposed to reduce its computational complexity, such as SC-Stack decoding (SCS) \cite{niu2012stack}, but at a cost of an increasing storage complexity.

A different approach has been proposed with SC-Flip (SCFlip) decoder, introduced in \cite{bastani2012polar} for the BEC channel, and later generalized to concatenated CRC-Polar codes over the Binary-Input Additive White Gaussian Noise (BI-AWGN) channel in \cite{afisiadis2014low}. The concept of SCFLip decoding is related to the ordered statistics decoding proposed in \cite{fossorier1995soft}, whose applicability to decoding short Polar and concatenated CRC-Polar codes has been recently investigated in \cite{wu2016ordered}. The principle is to allow a given number of new decoding attempts, in case that a failure of the initial SC decoding  is detected by the CRC. Each new decoding attempt consists in flipping one single hard decision bit -- starting with the least reliable one, according to the absolute value of the corresponding Log-Likelihood Ratio (LLR) -- of the initial SC decoding attempt, then decoding the subsequent positions by using the standard SC decoding. The above procedure is repeated until the CRC is verified or a predetermined maximum number of decoding attempts is reached. The SCFlip decoder provides a tunable trade-off between decoding performance and decoding complexity, since each new decoding attempt is only performed if the previous one failed. In particular, the average computational complexity of the SCFlip decoder tends to the one of the SC decoder at medium to high SNR, while competing with the CRC-aided SCL with list size $L = 2$, in terms of error correction performance \cite{afisiadis2014low}.   

In this work, we propose two improvements to the SCFlip decoding, based on refining and expanding some of the concepts we previously introduced in \cite{CSDimprovedSCFlip}. First, a new metric is proposed, aimed at determining the flipping positions that are more likely to {\em correct the trajectory} of the SCFlip decoding, {\em i.e.}, those positions that, once flipped, are more likely to lead to a successful decoding attempt.  
The proposed metric takes into account the sequential aspect of the SC decoder, and is shown to yield an improved error correction performance and a reduced computational complexity, as compared to the conventional LLR-based metric from \cite{afisiadis2014low}. Secondly, we introduce a generalization of the SCFlip decoder by considering not only one single bit-flip per new decoding attempt, but a number of $\omega \geq 1$ nested bit-flips.
These two improvements are materialized in a \textit{Dynamic SCFlip decoder} (D-SCFlip), in which the flipping positions are chosen dynamically by taking into consideration all the previous attempts, so that the next attempt is guaranteed to be the one with the best probability of success according to the optimized metric. The D-SCFlip decoder is shown to compete with the CRC-aided SCL decoder with list size  up to $L = 16$ in terms of decoding performance, while having an average computational complexity similar to that of the standard SC decoding at medium to high SNR. Moreover, we derive lower bounds on the Word Error Rate (WER) performance of any SCFlip decoder with the number of bit-flips per decoding attempt bounded by a maximum value $\omega$, and show that the D-SCFlip tightly approaches the WER lower bounds for $\omega\in\{1,2\}$.

The remainder of the paper is organized as follows. Section~\ref{sec:preliminaries} provides a short background on polar codes and main SC-based decoding algorithms. Section~\ref{sec:GeneralizedSCF} introduces the concept of bit-flips of order $\omega\geq 1$, and defines the general structure of a SCFlip decoder relying on higher-order bit-flips. Theoretical lower bounds on the WER performance of such a decoder are derived in Section~\ref{sec:idealSCF}.
Section~\ref{sec:optimized_metric} presents the proposed bit-flip metric, and investigates its efficiency in determining bit-flips leading to successful decoding attempts. The proposed D-SCFlip algorithm is finally described in Section~\ref{sec:dscflip}, where Monte-Carlo simulation results are also provided for performance evaluation  and comparison with other state of the art decoding techniques.



\section{Preliminaries}\label{sec:preliminaries}

\subsection{Polar Codes and Successive Cancellation Decoding}\label{subsec:scdecoder}

A Polar Code \cite{arikan2009channel} is characterized by a three-tuple $(N,K,\mathcal{I})$, where $N=2^n$ is the code-length, $K$ is the number of information bits, and $\mathcal{I} \subset \{1,...,N\}$ is a set indicating the positions of the $K$ information bits. Bits corresponding to positions $i\not\in {\cal I}$ are referred to as \textit{frozen bits} and are fixed to pre-determined values known at both the encoder and the decoder. 

We denote by $\textbf{U}=u_1^{N}$ the \textit{data vector}, of length $N$, containing $K$ information bits at  positions $i\in \mathcal{I}$, and $N-K$ frozen bits  at  positions $i\not\in \mathcal{I}$, which are assumed to be set to zero. The \textit{encoded vector}, denoted by $\mathbf{X}$, is obtained by:
$$\mathbf{X}=\textbf{U} \cdot \mathbf{G}_{N}$$
where $\mathbf{G}_N$ is the generator matrix \cite{arikan2009channel}. We further denote by $\mathbf{Y}$ the data received from the channel and used at the decoder input. $\mathbf{\hat{U}}=\hat{u}_1^{N}$ denotes the decoder's output, with $\hat{u}_i$ being the hard decision estimate of the bit $u_i$.

In SC decoding, each hard decision estimate $\hat{u}_i$ depends on both $\mathbf{Y}$ and the previous estimates $\hat{u}_1^{i-1}$, and is computed  according to the sign of the LLR: 
\begin{equation}
\text{L}_i=\log \left(\frac{\Pr(u_i=0 | \mathbf{Y}, \hat{u}_1^{i-1})}{\Pr(u_i=1 | \mathbf{Y}, \hat{u}_1^{i-1})} \right)
\end{equation}

\noindent by using the hard decision function $h$:
\begin{equation}
\label{eq:equation1}
\hat{u}_i = h(\text{L}_i) =  \left\{
    \begin{array}{ll}
        \qquad \displaystyle{u_i} & \mbox{if } i \notin \mathcal{I} \\
        \displaystyle{\frac{1 - \text{sign}(\text{L}_i)}{2}} & \mbox{if } i \in \mathcal{I}

    \end{array}
\right.
\end{equation}

\noindent where by convention $\text{sign}(0) = \pm 1$ with equal probability. 


\subsection{List decoding of Polar Codes}\label{subsec:scldecoder}
Due to its sequential nature, early errors occurring during the SC decoding process cannot be reversed. To overcome this problem, SCL decoding \cite{tal2011list} duplicates the SC decoding at each position $i\in {\cal I}$ in two parallel decoding threads, continuing in either possible direction.
In order to avoid an exponentially growing complexity, the number of parallel decoding paths is limited to a chosen, usually small, parameter $L$. The $L$ {\em surviving} decoding paths are determined according to a path metric, as discussed below. SCL decoder has a computational complexity growing as $\mathcal{O}(L \cdot N\log(N))$ and a space (memory) complexity of $\mathcal{O}(L \cdot N)$ \cite{tal2011list}. It has also been shown to closely approach the ML decoding performance if the size of the list $L$ is large enough. Moreover, in \cite{tal2011list} it has been observed that the SCL performance can be significantly improved, by concatenating an outer CRC code, to facilitate the identification of the correct decoding path among the list of $L$ candidates.

The SCL decoding computes a likelihood metric for each explored path, which can be alternatively expressed in the log-likelihood \cite{tal2011list}, or the log-likelihood ratio (LLR) \cite{DBLP:journals/corr/Balatsoukas-StimmingPB14} domain. In the LLR domain, the path metric is defined as follows:

\begin{definition}
For a path $l$ of length $i \leq N$, the path metric is defined by:
\begin{equation}\label{eq:path_metric}
PM[l]_i=\sum_{j=1}^{i} \log(1+\exp(-(1-2 \cdot \hat{u}[l]_j)\cdot \text{L}[l]_j))
\end{equation}
where
\begin{equation}
\text{L}[l]_j=\log \left(\frac{\Pr(u_j=0 | \mathbf{Y}, \hat{u}[l]_1^{j})}{\Pr(u_j=1 | \mathbf{Y}, \hat{u}[l]_1^{i})} \right)
\end{equation}
is the log-likelihood ratio of bit $u_j$ given the channel output $\mathbf{Y}$ and the past trajectory of the path $\hat{u}[l]_1^{j}$.
\end{definition}

\medskip\noindent 
Note that the sum in Eq.~(\ref{eq:path_metric}) is taken over all  $j=1,\dots,i$, including both frozen and non-frozen positions. However, for a frozen position $j\not\in{\cal I}$ the decoding path is not duplicated, thus $\hat{u}[l]_j = u_j$, irrespective of the $\text{L}[l]_j$ value. 

\noindent Several practical simplifications, aimed at reducing the computational complexity and/or the latency of the SCL decoding, as well as hardware implementations have been also proposed in the literature \cite{sarkis2013increasing, balatsoukas2014hardware, fan2015low}. An alternative to SCL decoding is the SCS decoding proposed in \cite{niu2012stack}, aimed at reducing the computational complexity, at a cost of a small loss in the error correction performance. Instead of exploring parallel decoding paths of the same length, the SCS uses an ordered stack of depth $D$, in which paths may have different lengths, and only the path with the largest path metric is extended. SCS decoding stops when the top path is of length $N$. The advantage of this decoder is that it is able to limit the number of operations compared to SCL decoder, especially when SC decoder is already able to decode correctly. The worst-case complexity of the SCS decoder is $\mathcal{O}(D \cdot N \log(N))$, but simulations show that the actual computational complexity is much lower, especially in moderate to high SNR regime. 


\section{Generalized SCFlip decoders}\label{sec:GeneralizedSCF}

Let $\mathcal{C}(N,K+r,\mathcal{I})$ denote the serial concatenation of an outer $(K+r, K)$ CRC code and an inner $(N, K+r, \mathcal{I})$ polar code. Note that the subset ${\cal I} \subset \{1, ..., N\}$ contains $K+r$ positions, for the $K$ information bits and the CRC of $r$ bits. 

The SCFlip decoder \cite{afisiadis2014low} consists of a standard SC decoding, possibly followed by a maximum number of $T$ new decoding attempts, until no errors are detected by the CRC check. Each new decoding consists of flipping one decision of the initial SC attempt, and decoding the subsequent positions by using the standard SC decoding. The position to be flipped is determined according to a given metric based on the LLRs obtained after the SC decoding. 

In this paper, we propose a generalization of the SCFlip decoder, by allowing a more than a single bit-flip for each decoding attempt. Therefore, we defined the notion of \textit{bit-flip of order $\omega$} as follows.
\begin{definition}\label{def:bit_flip_omega}
A \textit{(bit-)flip of order $\omega$} ($0 \leq \omega \leq K+r$) is a set of $\omega$ indices $\mathcal{E}=\{i_1, \dots i_{\omega} \} \subset \mathcal{I}$, such that $i_1< \dots < i_{\omega}$. The associated decoding attempt, denoted by SC$(\mathcal{E})$, corresponds to the SC decoding with the hard decision function $h$, defined in Eq. (2), replaced by $h_{\cal E}$, defined below:

\begin{equation}
\forall i \in \mathcal{I}, \quad \hat{u}[{\cal E}]_i = h_{\mathcal{E}}(\text{L}_i) \stackrel{\text{def}}{=} \left\{
    \begin{array}{ll}
        h(\text{L}_i) & \mbox{if } i \notin \mathcal{E} \\
        1-h(\text{L}_i) & \mbox{if } i \in \mathcal{E} 
    \end{array}
\right.
\end{equation}
This decoding attempt outputs a vector $\hat{u}[{\cal E}]_{1}^{N}$, the estimation of the codeword $u_1^{N}$. To simplify the notation, when no confusion is possible, $\hat{u}[{\cal E}]_{1}^{N}$ will be simply denoted by $\hat{u}_{1}^{N}$.
\end{definition}

Hence, in the \textit{Generalized SCFlip} (described in Algorithm~\ref{SCF1}), the decoding attempt associated to a bit-flip of order $\omega$ corresponds to a standard SC decision for each position, except for the the $\omega$ positions in $\mathcal{E}$, for which the decision is flipped. Note that if $\mathcal{E}=\emptyset$ (bit-flip of order $\omega=0$), the decoding attempt is exactly the same as the standard SC decoding. 

The exhaustive exploration of all the bit-flips of order $\omega \in \{0, ..., K+r\}$ would require a total number of $\displaystyle{\sum_{\omega=0}^{K+r} \dbinom{K+r}{\omega} = 2^{K+r}}$ decoding attempts, which is obviously too complex for a practical decoding solution. Therefore, we further equip the Generalized SCFlip decoder with a list $\mathcal{L}_{\text{flip}}=\{\mathcal{E}_1, \mathcal{E}_2, \dots \mathcal{E}_T\}$ of $T$ bit-flips of order $\omega_t, t \in \{1, \cdots T \}$.


\begin{algorithm}[!t]
\caption{Generalized SCFlip decoder}
\label{SCF1}
\begin{algorithmic}[1]
\Procedure{Generalized SCFlip}{$\mathbf{Y},\mathcal{I},T,\mathcal{L}_{\text{flip}}=\{\mathcal{E}_1, \dots,\mathcal{E}_{T} \}$}
	\State $\hat{u}_1^{N} \leftarrow $SC$(\emptyset)$ 	
	\IIf{CRC($\hat{u}_1^{N}$) = success} return $\hat{u}_1^{N}$; \EndIIf
	\For {$t=1,\dots,T$}	
	 	\State $\hat{u}_1^{N} \leftarrow $SC$(\mathcal{E}_t$); 
		\IIf{CRC($\hat{u}_1^{N}$) = success} return $\hat{u}_1^{N}$; \EndIIf 
	 \EndFor
	\State return $\hat{u}_1^{N}$;
\EndProcedure
\end{algorithmic}
\end{algorithm}

The Generalized SCFlip algorithm proceeds to at most $T+1$ decoding attempts, starting with the standard SC decoding and, followed by the decoding attempts SC(${\cal E}_t$), with ${\cal E}_t \in \mathcal{L}_{\text{flip}}$. The decoding process stops if:
\begin{itemize}
    \item one of the decoding attempt verifies the CRC
    \item all $T$ bit-flips from the list have been tested
\end{itemize}

The Generalized SCFlip decoder may recover the correct codeword, only if $\mathcal{L}_{\text{flip}}$ contains the unique bit-flip of order $\omega \geq 1$ that corrects the SC decoding trajectory (assuming that the initial SC decoding attempt failed).
However, having the correct bit-flip in $\mathcal{L}_{\text{flip}}$ does not guarantee successful decoding, since an earlier, erroneous decoding attempt might verify the CRC (undetected error), so that the decoding process stops and the following bit-flips in $\mathcal{L}_{\text{flip}}$ are not tested. 
The probability this happens depends on both the probability of undetected error of the CRC and the position of the 
correct bit-flip within $\mathcal{L}_{\text{flip}}$.  

In view of the previous discussion, the effectiveness of the Generalized SCFlip decoder depends directly on the way the list of tested bit-flips $\mathcal{L}_{\text{flip}}$ is determined. It also appears that rather than a predetermined list, $\mathcal{L}_{\text{flip}}$ should actually depend on the current noise realization, so as to
increase the probability of including the correct bit-flip ({\em i.e.}, correcting the SC decoding trajectory) in front positions.  Moreover, the information gathered during the decoding process (e.g., LLR values computed during the initial SC decoding or the following decoding attempts) can also be used to determine those bit-flips that are most likely to correct a given noise realization, and thus to dynamically update the list $\mathcal{L}_{\text{flip}}$. To do so, the candidate bit-flips have to be evaluated by a metric that estimates their likelihood to correct a given noise realization, which will be discussed in Section~ \ref{sec:optimized_metric}.

Before discussing the optimization of such a metric and the method to dynamically generate the list of bit-flips, in the next section we  derive lower bounds on the WER performance of the Generalized SCFlip decoder using bit-flips of order $\leq \omega$. 
These lower bounds will also serve as a reference for assessing the effectiveness of the proposed D-SCFlip decoder in Section~\ref{sec:dscflip}, and implicitly of the bit-flip metric proposed in Section~\ref{sec:optimized_metric}.


\section{Word Error Rate Lower Bound for Generalized SCFlip Decoders}\label{sec:idealSCF}

\subsection{Order of a noise realization} \label{subsec:order_noise_realization}

In the following, we shall use the expression {\em noise realization} to refer to the channel noise that corrupted the actually observed signal $\mathbf{Y}$. We say that a {\em noise realization is of order} $\omega$, if there exits a bit-flip $\mathcal{E}$ of order $\omega$, such that the observed signal $\mathbf{Y}$ is corrected by the SC($\mathcal{E}$) decoding attempt (see Definition 2). 
The order of a noise realization can be efficiently computed by using the Oracle-Assisted SC (OA-SC) decoder proposed in \cite{afisiadis2014low}. OA-SC performs the same operations as the standard SC decoder, but instead of propagating the hard decision estimates of the previous decoded bits, and thus risking to propagate an erroneous decision, it is helped by an oracle to propagate the correct decisions. Hence, the oracle-assisted LLR of the bit $u_i$, denoted by $\text{L}^{\text{OA}}_i$, can be expressed as:

\begin{equation}
\text{L}^{\text{OA}}_i=\log \left(\frac{\Pr(u_i=0 | \mathbf{Y}, u_1^{i})}{\Pr(u_i=1 | \mathbf{Y}, u_1^{i})} \right)
\end{equation}

\noindent and the hard decision estimate of $u_i$ is given by $\hat{u}_i^{\text{OA}}=h(\text{L}^{\text{OA}}_i)$. Let $\mathcal{E}_{\mathbf{Y}}=\{\text{ }i \in \mathcal{I} \text{  }|\text{  }\hat{u}_i^{\text{OA}} \ne u_i \}$ and $\omega_{\mathbf{Y}}=|\mathcal{E}_{\mathbf{Y}}|$ be the order ({\em i.e.} number of elements) of $\mathcal{E}_{\mathbf{Y}}$. Then the order of the noise realization is equal to $\omega_{\mathbf{Y}}$, and the observed signal $\mathbf{Y}$ is successfully corrected by the SC($\mathcal{E}_{\mathbf{Y}}$) decoding.  

\subsection{WER Lower Bound}\label{subsec:wer_lower_bounds}

Let SCFlip-$\omega$ denote a Generalized SCFlip decoder (Algorithm \ref{SCF1}) whose maximum bit-flip order is equal to $\omega$. Hence, using the notation from Section III, $\omega= \displaystyle{\max_{\mathcal{E} \in \mathcal{L}_{\text{flip}}}} |\mathcal{E}|$. Such a decoder successfully corrects a noise realization of order $\omega_{\mathbf{Y}} \leq \omega$ if and only if (i) the corresponding bit-flip $\mathcal{E}_{\mathbf{Y}} \in \mathcal{L}_{\text{flip}}$ and (ii) no previous decoding attempt SC($\mathcal{E}$) satisfies the CRC check before SC($\mathcal{E}_{\mathbf{Y}}$). 
We further denote by $\mathtt{i}$SCFlip-$\omega$ the {\em ideal} SCFlip-$\omega$ decoder that successfully corrects any noise realization of order less than or equal to $\omega$. The ideal $\mathtt{i}$SCFlip-$\omega$ decoder can be seen as an SCFlip-$\omega$ decoder such that (i) $\mathcal{L}_{\text{flip}}$ contains all the bit-flips of order less than or equal to $\omega$, hence the list size is given by $T=\displaystyle{\sum_{\omega^{'}=1}^{\omega}} \dbinom{K+r}{\omega^{'}}$, and (ii) the CRC error detection is replaced by an ideal detector, which is satisfied only for the correct word. 

The WER of any SCFlip-$\omega$ is lower-bounded by the WER of the $\mathtt{i}$SCFlip-$\omega$ decoder. The latter can be efficiently determined by running the OA-SC decoder to compute the order $\omega_{\mathbf{Y}}$ of the actual noise realization, then declaring a decoding failure if and only if $\omega_{\mathbf{Y}} > \omega$. It is worth noticing that this lower bound, further referred to as the {\em ideal WER of order} $\omega$ ($\mathtt{i}$WER-$\omega$), is not necessarily achievable by the SCFlip-$\omega$ decoder and the $\mathtt{i}$WER-$\omega$ lower bound can even be better than the ML performance in some cases (an $\mathtt{i}$SCFlip-$\omega$ decoder with $\omega=K+r$ would be able to correct any noise realization). However, for small $\omega$ values, the ideal WER can be closely approached by practical SCFlip-$\omega$ decoders, provided that the CRC is reliable enough, as it will be shown in Sections \ref{sec:optimized_metric}-\ref{sec:dscflip}.

\begin{figure}[t!]
    \centering
    \includegraphics[width=0.6\linewidth]{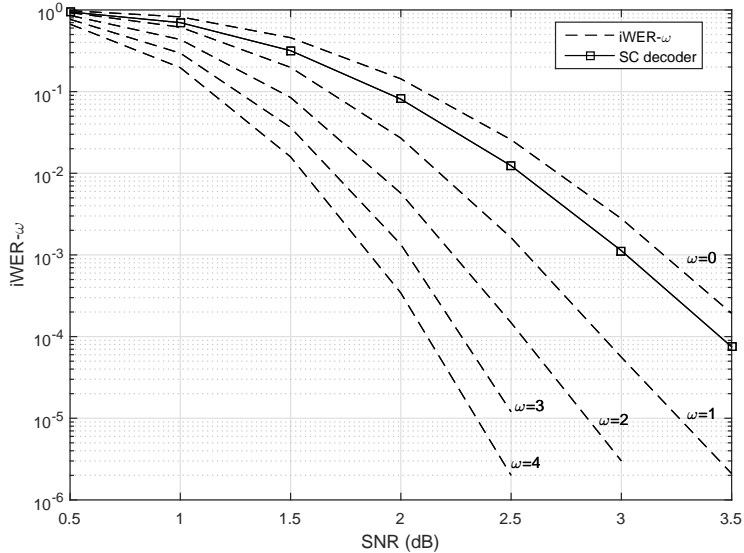}
    \caption{Performance of ideal SCFlip-$\omega$ decoder for a code $(N,K+r)=(1024,512+16)$}
    \label{fig:iwer_omega}
\end{figure}


Figure \ref{fig:iwer_omega} presents the lower-bounds of SCFlip-$\omega$ decoders with $\omega=\{0,1,2,3,4\}$ for a CRC-concatenated polar code with parameters $(N,K+r)=(1024,512+16)$. For $\omega = 0$,  $\mathtt{i}$WER-$0$ corresponds to  the WER of the SC decoder for a polar code of length $N$, with $K+r$ information bits. 
Moreover, we also plot the performance of the SC decoder with $(N,K)=(1024,512)$, which is better than the $\mathtt{i}$SCFlip-$0$ performance, due to the higher number of frozen bits.
It can be seen that $\mathtt{i}$SCFlip-$\omega$ decoders exhibit significant SNR gains compared to the SC decoder,  from $0.5$\,dB for the $\mathtt{i}$SCFlip-1, to about $1$\,dB for the $\mathtt{i}$SCFlip-2 decoder, at WER $=10^{-4}$. 

\subsection{Impact of the code-length and coding-rate on the ideal WER}\label{subsec:impact_on_iWER}

This section investigates the ideal decoding performance of $\mathtt{i}$SCFlip-$\omega$ decoders, for various code-lengths and coding rates, and small values of $\omega$. More precisely, we investigate the relation between $\mathtt{i}\text{WER}\omega$ for $\omega=\{1,2,3\}$ and $\mathtt{i}\text{WER-}0$, as function of the coding rate $R=\frac{K}{N}$ and code-length $N$:
\begin{equation}\label{eq:iweromega_iwer0}
\mathtt{i}\text{WER}\omega =f^{(\omega)}_{N,R}(\mathtt{i}\text{WER-}0)
\end{equation}

\begin{figure}[t!]
    \centering
    \subfigure[\vspace*{-5mm}$N = 1024$, varying $R \in \{1/3, 1/2, 2/3 \}$]{\includegraphics[width=0.48\linewidth]{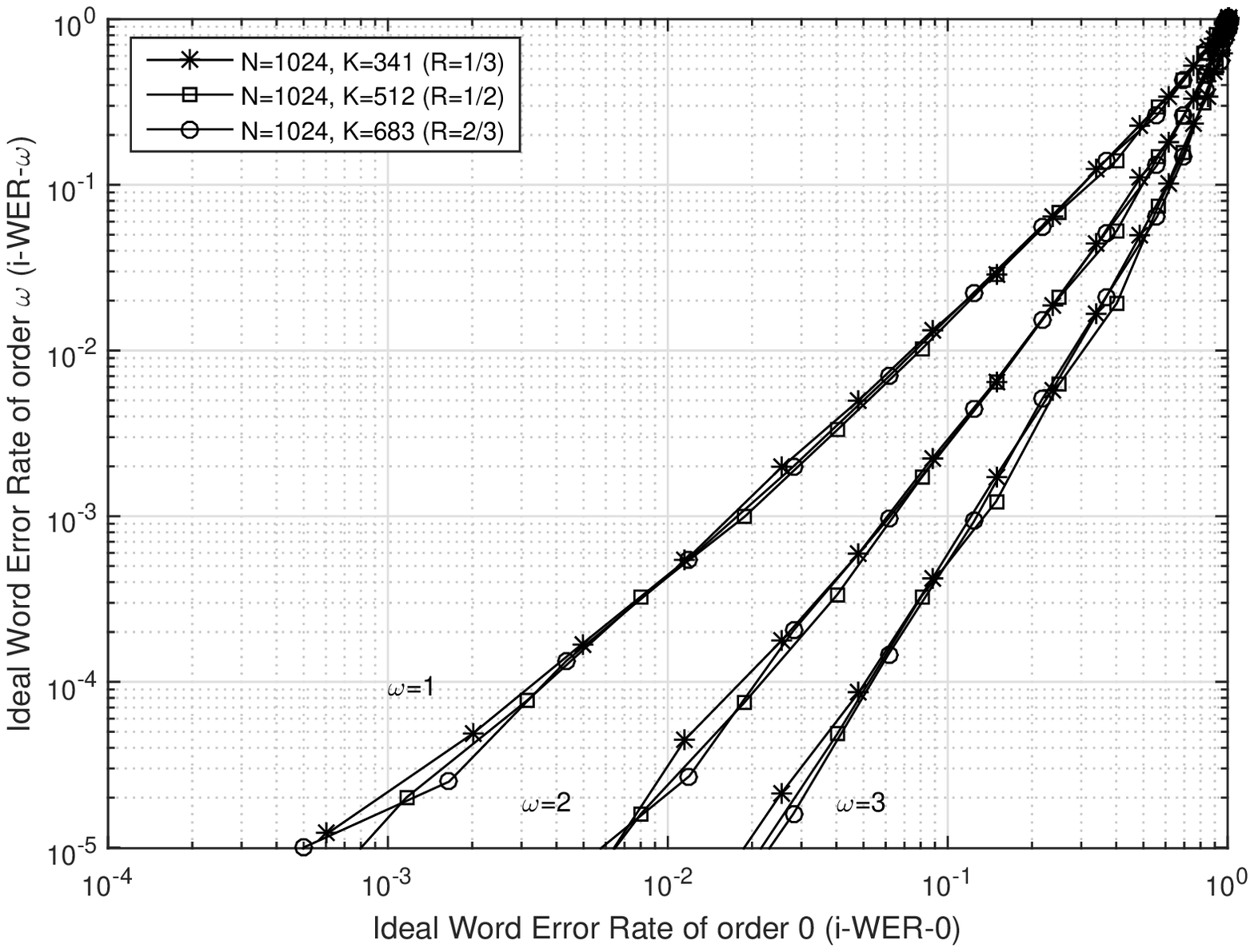}\label{fig:iwer_length_cst}}%
\hfill\subfigure[$R = 1/2$ and varying $N \in \{512,1024,2048 \}$]{\includegraphics[width=0.48\linewidth]{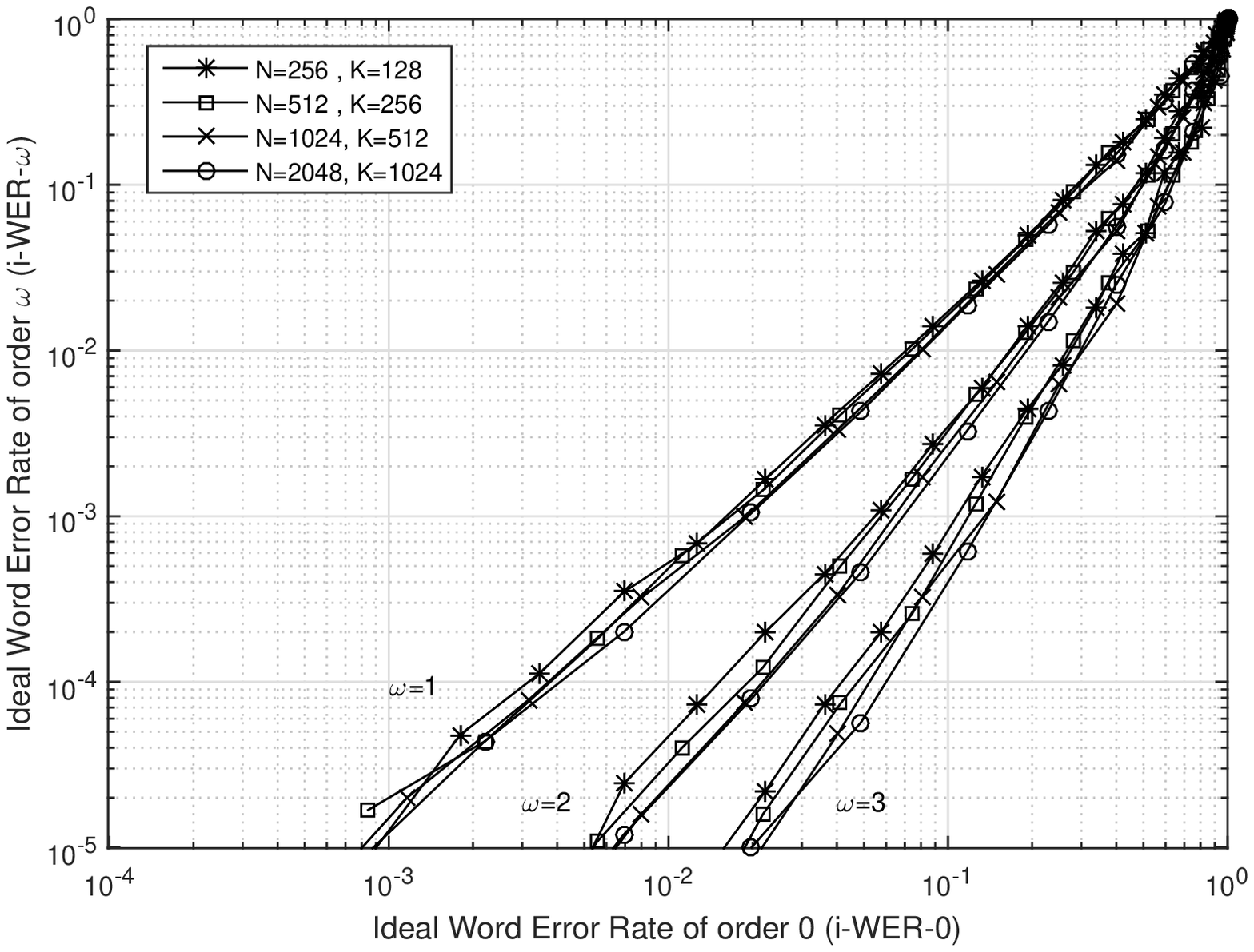}\label{fig:iwer_rate_cst}}
\vspace*{5mm}    
    \caption{$\mathtt{i}$WER-${\omega}$ as function of $\mathtt{i}\text{WER}$-$0$ for varying coding rate $R$ or varying code length $N$}
\end{figure}


The study is divided into two parts. (a) First, for a given code-length $N$, we observe this function for different coding rates $R$. Figure~\ref{fig:iwer_length_cst} plots $\mathtt{i}\text{WER}\omega$, for $\omega \in \{1,2,3\}$, as a function of $\mathtt{i}\text{WER-}0$ (assuming BI-AWGN channel), for a code-length $N=1024$ and coding rates $R\in\{1/3, 1/2, 2/3 \}$. It can be observed that for a given value of $\omega$, the $\mathtt{i}\text{WER}\omega$ depends only on $\mathtt{i}\text{WER}\text{-}0$ and is practically independent of the coding rate $R$.
 (b) Second, as shown in Fig.~\ref{fig:iwer_rate_cst}, a similar observation can be made if one considers a fixed coding rate $R=1/2$, and variable code-length $N$. Therefore, we conclude that $\mathtt{i}\text{WER}\omega$ essentially depends on $\mathtt{i}\text{WER}\text{-}0$, and thus Eq.~(\ref{eq:iweromega_iwer0}) can be approximated to:
 \begin{equation}
    \mathtt{i}\text{WER}\omega \simeq f^{(\omega)}(\mathtt{i}\text{WER}\text{-}0)
\end{equation}

Since $\mathtt{i}\text{WER}\omega = 1-\Pr(\omega_{\mathbf{Y}} \leq \omega)$, this analysis translates into the following interesting property: consider two polar codes $C_1(N_1,K_1)$ and $C_2(N_2,K_2)$, with the same WER performance under SC decoding, at SNR$_1$ and SNR$_2$, respectively.  Then the noise realization orders $\omega_{\mathbf{Y}_1}$ and $\omega_{\mathbf{Y}_2}$ are expected to follow nearly the same probability distribution.


From a practical point of view, this analysis can also be used to determine different sets of code and decoder parameters that would be able to achieve a target WER performance (assuming a given SNR). Indeed, $\mathtt{i}\text{WER}\text{-}0$ can be easily estimated, {\em e.g.}, by using the density evolution technique \cite{mori2009performance, trifonov2012efficient}. Hence, in order to achieve a target WER, for example of $10^{-4}$, the  code  parameters $(N,K+r)$ must be chosen such that $\mathtt{i}\text{WER}\text{-}0 \approx 3 \cdot 10^{-3}$ for an SCFlip-1 decoder, or such that $\mathtt{i}\text{WER}\text{-}0 \approx3 \cdot 10^{-2}$ for an SCFlip-2 decoder. Of course, the SCFlip-1 and SCFlip-2 decoders under use should be able to closely approach the corresponding lower bounds, $\mathtt{i}\text{WER-1}$ and $\mathtt{i}\text{WER-2}$. In the following section, we will show that these lower bounds can be indeed tightly approached by practical decoders. 

\section{Optimized Metric for Generalized SCFlip Decoders}\label{sec:optimized_metric}

In order to build practical SCFlip-$\omega$ decoders that closely approach the  ideal performance of $\mathtt{i}$SCFlip-$\omega$, we first introduce an optimized bit-flip metric, adapted to bit-flips of any order $\omega \geq 1$, then we propose an efficient strategy to build the bit-flips list $\mathcal{L}_{\text{flip}}$. In this section we describe the proposed metric, while the construction of $\mathcal{L}_{\text{flip}}$ will be discussed in the next section.

\subsection{Proposed Metric for Generalized SCFlip Decoder}\label{subsec:proposed_metric}

The SCFlip decoder from \cite{afisiadis2014low} considers only bit-flips of order $1$, which are chosen according to the absolute value of the corresponding LLR. Thus, in case the initial SC decoding fails, the selected bit-flips of order $1$ correspond to the $T$ positions $i \in {\cal I}$ with the lowest $|L_i|$ values. However,  using the absolute value of the LLR as likelihood metric for a bit-flip is sub-optimal, since it does not take into account the sequential aspect of the SC decoder. Indeed, while a lower absolute value of the LLR indicates that the corresponding hard decision bit has a higher error probability, it does not provide any information about the probability of being the first error that occurred during the sequential decoding process. In other words, such a metric does not distinguish the very first error from the subsequent ones.

We propose a new metric, aimed at evaluating the likelihood of a bit-flip $\mathcal{E}_{\omega}=\{i_1,\dots,i_{\omega} \} \subset \mathcal{I}$, of order $\omega$, to correct the trajectory of the SC decoding. By {\em correcting the trajectory of the SC decoding}, we mean that SC($\mathcal{E}_{\omega}$) successfully decodes all the bits $u_i$ with $i \leq i_{\omega}$ (recall that indices $i_1,\dots,i_{\omega}$ are assumed to be in increasing order).  Note that this does not mean that the SC($\mathcal{E}_{\omega}$) decoding is successful, since there is no guarantee that it will successfully decode the subsequent bits, {\em i.e.}, bits $u_i$ with $i>i_{\omega}$. For instance, for $\omega=1, \mathcal{E}_1=\{i_1\}$ corrects the trajectory of the SC decoding if and only if $i_1$ is the first erroneous position of the SC decoding attempt, but this does not guarantee that the SC($\mathcal{E}_1$) is successful. 

For any $1 \leq \omega' \leq \omega$, let $\mathcal{E}_{\omega'}=\{i_1,\dots,i_{\omega'} \}$ be the bit-flip of order $\omega'$ determined by the first $\omega'$ indices in $\mathcal{E}_{\omega}$.
Let $\text{L}[\mathcal{E}_{\omega'}]_i$, $\hat{u}[\mathcal{E}_{\omega'}]_i$ denote respectively the LLR and the hard decision estimate computed by SC($\mathcal{E}_{\omega'}$), corresponding to bit $u_i$. 
According to Definition~\ref{def:bit_flip_omega}, SC($\mathcal{E}_{\omega'}$) and SC($\mathcal{E}_{\omega'-1}$) are identical for positions $i<i_{\omega'}$, while for $i=i_{\omega'}$, SC($\mathcal{E}_{\omega'}$) flips the hard decision estimate computed by SC($\mathcal{E}_{\omega'-1}$). Hence, for any $\omega' \leq \omega$, one has:
\begin{equation}\label{eq:Lomega_recursion}
\text{L}[\mathcal{E}_{\omega'}]_i=\text{L}[\mathcal{E}_{\omega'-1}]_i   \qquad \forall i \leq i_{\omega^{'}} 
\end{equation}
\begin{equation}\label{eq:Uomega_recursion}
\hat{u}[\mathcal{E}_{\omega'}]_i=\hat{u}[\mathcal{E}_{\omega'-1}]_i,   \ \forall i < i_{\omega'} \ \text{   and   }\  \hat{u}[\mathcal{E}_{\omega'}]_{i_{\omega'}}=1-\hat{u}[\mathcal{E}_{\omega'-1}]_{i_{\omega'}}
\end{equation}

Let $P(\mathcal{E}_{\omega})$ denote the probability of $\mathcal{E}_{\omega}$ correcting the trajectory of SC. It follows that:
\begin{align}
\begin{split}
  P(\mathcal{E}_{\omega}) & = \text{Pr}(\hat{u}[\mathcal{E}_{\omega}]_1^{i_{\omega}}=u_1^{i_{\omega}}  | \mathbf{Y})\\
  &=\text{Pr}(\hat{u}[\mathcal{E}_{\omega-1}]_{i_{\omega}} \ne u_{i_{\omega}} ,\hat{u}[\mathcal{E}_{\omega-1}]_1^{i_{\omega-1}}=u_1^{i_{\omega-1}}  | \mathbf{Y})\\
  &= p_{e}(\hat{u}[\mathcal{E}_{\omega-1}]_{i_{\omega}})\cdot \prod_{j=i_{\omega-1}+1}^{i_{\omega}-1} (1-p_{e}(\hat{u}[\mathcal{E}_{\omega-1}]_j))  \cdot P(\mathcal{E}_{\omega-1}) \\
\end{split}
\end{align}
where $p_{e}(\hat{u}[\mathcal{E}_{\omega-1}]_{j})\overset{\text{def}}{=}\text{Pr}\left(\hat{u}[\mathcal{E}_{\omega-1}]_j \ne u_j |  \mathbf{Y}, \hat{u}[\mathcal{E}_{\omega-1}]_1^{j}=u_1^{j}\right)$. By taking into account Eq.~(\ref{eq:Uomega_recursion}), the above recursion can be unfolded to the following expression:
\begin{equation}\label{eq:proba_Eomega}
P(\mathcal{E}_{\omega}) = 
    \prod_{j \in \mathcal{E}_{\omega}}  p_{e}(\hat{u}[\mathcal{E}_{\omega-1}]_{j}) 
    \ \cdot 
    \prod_{\underset{j \in \mathcal{I} \setminus \mathcal{E}_{\omega}}{j<i_{\omega}}} 
    \left( 1-p_{e}(\hat{u}[\mathcal{E}_{\omega-1}]_j)  \right)
\end{equation}

\noindent Note that the second product on the right-hand side term of Eq.~(\ref{eq:proba_Eomega}) is taken only over indexes $j\in {\cal I}$, since $p_{e}(\hat{u}[\mathcal{E}_{\omega-1}]_{j})=0$ for  $j \notin \mathcal{I}$. Computing $p_{e}(\hat{u}[\mathcal{E}_{\omega-1}]_{j})$ is an arduous task, since this probability is conditional on the fact that the previous bits have been correctly decoded by SC($\mathcal{E}_{\omega-1}$).  
Instead, one can compute the probability $q_{e}(\hat{u}[\mathcal{E}_{\omega-1}]_{j}) \overset{\text{def}}{=} \text{Pr}(\hat{u}[\mathcal{E}_{\omega-1}]_j \ne u_j |  \mathbf{Y}, \hat{u}[\mathcal{E}_{\omega-1}]_1^{j})$, which is conditional on the previously decoded bits, irrespective of whether they have been correctly decoded or not, and is given by (this follows directly from the definition of $\text{L}[\mathcal{E}_{\omega-1}]_j$ and $\hat{u}[\mathcal{E}_{\omega-1}]_{j}$):
\begin{equation}
 q_{e}(\hat{u}[\mathcal{E}_{\omega-1}]_{j}) = 
 \frac{1}{1+\exp{(|\text{L}[\mathcal{E}_{\omega-1}]_{j}|)}}, \qquad\forall j\in {\cal I} 
\end{equation}

Hence, we propose to use $q_{e}(\hat{u}[\mathcal{E}_{\omega-1}]_{j})$ as an approximation of $p_{e}(\hat{u}[\mathcal{E}_{\omega-1}]_{j})$, and we further introduce a parameter $\alpha$ (see below) as a mean to compensate this approximation. In practice, the value of $\alpha$ can be optimized by Monte-Carlo simulation, as shown in Section~\ref{subsec:optimal_alpha}. 
Using $p_{e}(\hat{u}[\mathcal{E}_{\omega-1}]_{j}) \approx \frac{1}{1+\exp{(\alpha|\text{L}[\mathcal{E}_{\omega-1}]_{j}|)}}$ in Eq.~(\ref{eq:proba_Eomega}), we obtain the following metric, denoted by $M_{\alpha}(\mathcal{E}_{\omega})$, which will be used to approximate the probability of $\mathcal{E}_{\omega}$ correcting the trajectory of SC:

\begin{definition}
The metric associated with a bit-flip $\mathcal{E}_{\omega}= \{i_1,\dots,i_{\omega}\} \subset \mathcal{I}$, of order $\omega$, is defined by:
\begin{equation}
\label{eq:equation4}
M_{\alpha}(\mathcal{E}_{\omega})=\prod_{j \in \mathcal{E}_{\omega}} \left( \frac{1}{1+\exp{(\alpha|\text{L}[\mathcal{E}_{\omega-1}]_{j})|)}} \right) \cdot \prod_{\underset{j \in \mathcal{I} \setminus \mathcal{E}_{\omega}}{j<i_{\omega}}} \left( \frac{1}{1+\exp{(-\alpha|\text{L}[\mathcal{E}_{\omega-1}]_{j}|)}} \right)
\end{equation}
\end{definition}

\noindent Note that for a bit flip $\mathcal{E}_1=\{i_1\}$ of order $1$, the above metric can be written as:
\begin{equation}\label{eq:metric_order1}
M_{\alpha}(\mathcal{E}_1)= \frac{1}{1+\exp{(\alpha|\text{L}_{i_1}|)}} \cdot \prod_{\underset{j \in \mathcal{I}}{j<i_1}} \left( \frac{1}{1+\exp{(-\alpha|\text{L}_{j}|)}} \right),
\end{equation}
where $\text{L}_j$ are the LLR values computed by the initial SC decoding attempt. Moreover, the metric of the bit-flip $\mathcal{E}_{\omega}$ can be computed recursively, using the following equation:
\begin{equation}
\label{eq:equation3}
M_{\alpha}(\mathcal{E}_{\omega})=\frac{1}{1+\exp{(\alpha |\text{L}[\mathcal{E}_{\omega-1}]_{i_{\omega}}|)}} \cdot \prod_{\underset{j \in \mathcal{I}}{j=i_{\omega-1}+1}}^{i_{\omega}-1} \left( \frac{1}{1+\exp{(-\alpha |\text{L}[\mathcal{E}_{\omega-1}]_{i_{\omega}}|)}} \right) \cdot M_{\alpha}(\mathcal{E}_{\omega-1})
\end{equation}
Indeed, by taking into account Eq.~(\ref{eq:Lomega_recursion}), it can be easily seen that the above recursion unfolds to the expression from Eq.~(\ref{eq:equation4}).


Using the fact that $\frac{1}{1+\exp(x)}=\frac{\exp(-x)}{1+\exp(-x)}$,  Eq.~(\ref{eq:equation4}) can  be rewritten:
\begin{equation}\label{eq:metric_rewritten}
M_{\alpha}(\mathcal{E}_{\omega})=\prod_{j \in \mathcal{E}_{\omega}} \exp{(-\alpha|\text{L}[\mathcal{E}_{\omega-1}]_{j})|)} \cdot \prod_{\underset{j \in \mathcal{I}}{j \leq i_{\omega}}} \left( \frac{1}{1+\exp{(-\alpha|\text{L}[\mathcal{E}_{\omega-1}]_j|)}} \right)
\end{equation}

\noindent By taking the logarithm of this formula, and denoting $M'_{\alpha}(\mathcal{E}_{\omega})=-\frac{1}{\alpha} \cdot \log(M_{\alpha}(\mathcal{E}_{\omega}))$, one gets the following equivalent  metric in the logarithmic domain:
\begin{align}
\label{eq:equation5}
\begin{split}
  M'_{\alpha}(\mathcal{E_{\omega}}) &  = \sum_{j \in \mathcal{E_{\omega}}}|\text{L}[\mathcal{E}_{\omega-1}]_j| + S_{\alpha}(\mathcal{E}_{\omega}) \\
  \text{where } S_{\alpha}(\mathcal{E}_{\omega}) & = \frac{1}{\alpha} \sum_{\underset{j \in \mathcal{I}}{j \leq i_{\omega}}} \log( 1+\exp(-\alpha \cdot |\text{L}[\mathcal{E}_{\omega-1}]_j|))
\end{split}
\end{align}


On the basis of the above considerations,  the list $\mathcal{L}_{\text{flip}}$ used within a generalized SCFlip-$\omega$ decoder should be constituted of bit-flips with the highest probability-domain metric $M_{\alpha}$, or equivalently with the lowest logarithmic domain metric $M_{\alpha}'$, since they are the most likely to correct the trajectory of the SC decoding. For the sake of simplicity, the algorithms proposed in the next sections will be defined by using the metric $M_{\alpha}$, but it is worth mentioning that the logarithm domain metric $M'_{\alpha}$ is more suitable for practical implementations, due to its better numerical stability. 

\subsection{Impact of the $\alpha$ parameter}\label{subsec:impact_alpha}

In order to understand the impact of the parameter $\alpha$ on the proposed metric, we start by considering two limiting cases, namely $\alpha = 0$ and $\alpha \rightarrow +\infty$. 

For $\alpha = 0$, using Eq.~(\ref{eq:metric_rewritten}), it can be seen that  $M_0(\mathcal{E}_\omega) = \left(\frac{1}{2}\right)^{k_{\cal I}}$, where $k_{\cal I}$ is the number of positions in ${\cal I}$ less than or equal to $i_{\omega}$. 
Therefore, if $\mathcal{E}_\omega = \{i_1,\dots,i_{\omega}\}$ and $\mathcal{E}'_{\omega'}= \{i'_1,\dots,i'_{\omega'}\}$ are two bit-flips of order $\omega$ and $\omega'$, $M_0(\mathcal{E}_\omega) \geq M_0(\mathcal{E'}_{\omega'}) \Leftrightarrow i_\omega \leq i'_{\omega'}$. In other words, bit-flips are ordered by $M_0$ according to the index of their last flipped position. 

For $\alpha \rightarrow +\infty$, we consider the equivalent logarithmic-domain metric defined in Eq.~(\ref{eq:equation5}). It can be seen that $\lim\limits_{\alpha \rightarrow +\infty} S_{\alpha}(\mathcal{E}_\omega)=0$, thus $M'_{\infty}(\mathcal{E}_\omega) = \sum_{j \in \mathcal{E_{\omega}}}|\text{L}[\mathcal{E}_{\omega-1}]_j|$ is the sum of {\em reliabilities} ({\em i.e.} absolute value of the LLR) of the flipped positions. In the particular case of bit-flips of order $1$, this metric is exactly the same as the one in \cite{afisiadis2014low}.

In general, for $0 < \alpha < +\infty$, $S_{\alpha}(\mathcal{E}_\omega)$ can be seen as a penalty added to $\sum_{j \in \mathcal{E_{\omega}}}|\text{L}[\mathcal{E}_{\omega-1}]_j|$, which takes into consideration the sequential aspect of the SC decoding, providing and intermediate and tunable solution between prioritizing bit-flips according to either the index of their last flipped position or the sum of reliabilities of the flipped positions.

The value of the trade-off parameter $\alpha$ can be optimized by Monte-Carlo simulation. It is expected that the optimal $\alpha$ value decreases with the SNR. Indeed, considering Eq.~(\ref{eq:equation5}) for a fixed $\alpha$ value, and taking the limit as the SNR goes to infinity, the term $S_{\alpha}(\mathcal{E}_\omega)$ tends to $0$ and becomes negligible compared to $\sum_{j \in \mathcal{E_{\omega}}}|\text{L}[\mathcal{E}_{\omega-1}]_j|$, and therefore the sequential characteristic of the decoder is no longer accounted for by the considered metric.  
Consequently, it is expected that the optimal value of $\alpha$ will decrease with the SNR, so that to rebalance the contribution of the $S_{\alpha}(\mathcal{E}_\omega)$ term to the value of the considered metric. This is confirmed by the Monte-Carlo simulations presented in section \ref{subsec:optimal_alpha}. 

Finally, it is worth underlining the strong similarity between the derived bit-flip metric and the path metric used by the SCL decoder (Eq.~(\ref{eq:path_metric})). However, unlike the SCL path metric, frozen bits do not contribute to our proposed bit-flip metric.


\subsection{Optimization of the $\alpha$ parameter}\label{subsec:optimal_alpha}

This section investigates the optimization of the $\alpha$ parameter, so that to increase the probability that the bit-flip $\mathcal{E}_{\mathbf{Y}}$ is ranked high by the metric $M_\alpha$, where $\mathcal{E}_{\mathbf{Y}}$  is bit-flip correcting the SC decoding trajectory, for the the current noise realization $\mathbf{Y}$ (see Section~\ref{subsec:order_noise_realization}).


Let $\bar{\cal L}_{\alpha,\mathbf{Y}}$ denote the list of all the bit-flips ${\cal E}$, of any order $\omega = 1,\dots, K+r$, ordered according to decreasing values of $M_\alpha({\cal E})$. Note that the bit-flips ordering depends on both the value of  $\alpha$  and the current noise realization $\mathbf{Y}$.  
We denote by $\text{rk}_\alpha(\mathcal{E}_{\mathbf{Y}})$ the rank (position) of $\mathcal{E}_{\mathbf{Y}}$ within the ordered list $\bar{\cal L}_{\alpha,\mathbf{Y}}$. Let $\mathcal{E}_{\mathbf{Y}} = \{ i_1, \dots, i_{\omega_{\mathbf{Y}}} \}$, where $\omega_{\mathbf{Y}}\geq 1$ is the order of $\mathcal{E}_{\mathbf{Y}}$. Using the recursion from Eq.~(\ref{eq:equation3}), it follows that:
\begin{equation}
M_\alpha(\{ i_1 \}) > M_\alpha(\{ i_1, i_2 \}) > \cdots > M_\alpha(\{ i_1, \dots, i_{\omega_{\mathbf{Y}}} \})
\end{equation}
and therefore:
\begin{equation}
\text{rk}_\alpha(\mathcal{E}_{\mathbf{Y}}) \geq \omega_{\mathbf{Y}}
\end{equation}
Finally, we define the optimal $\alpha$ value, denote by $\alpha_{\text{opt}}$, as:

\begin{equation}
\alpha_{\text{opt}} = \mathop{\text{argmin}}_\alpha \mathbb{E}\left(\text{rk}_\alpha(\mathcal{E}_{\mathbf{Y}})\right),
\end{equation} 
where $\mathbb{E}\left(\text{rk}_\alpha(\mathcal{E}_{\mathbf{Y}})\right)$ denotes the expected value of the random variable $\text{rk}_\alpha(\mathcal{E}_{\mathbf{Y}})$, assuming that $\omega_{\mathbf{Y}} \geq 1$ ({\em i.e.}, SC  fails to decode the current noise realization $\mathbf{Y}$).

\begin{figure}[t!]
    \centering
    \subfigure[\vspace*{-5mm}$\alpha_{\text{opt}}$ as function of $\mathtt{i}$WER-$0$]{\includegraphics[width=0.48\linewidth]{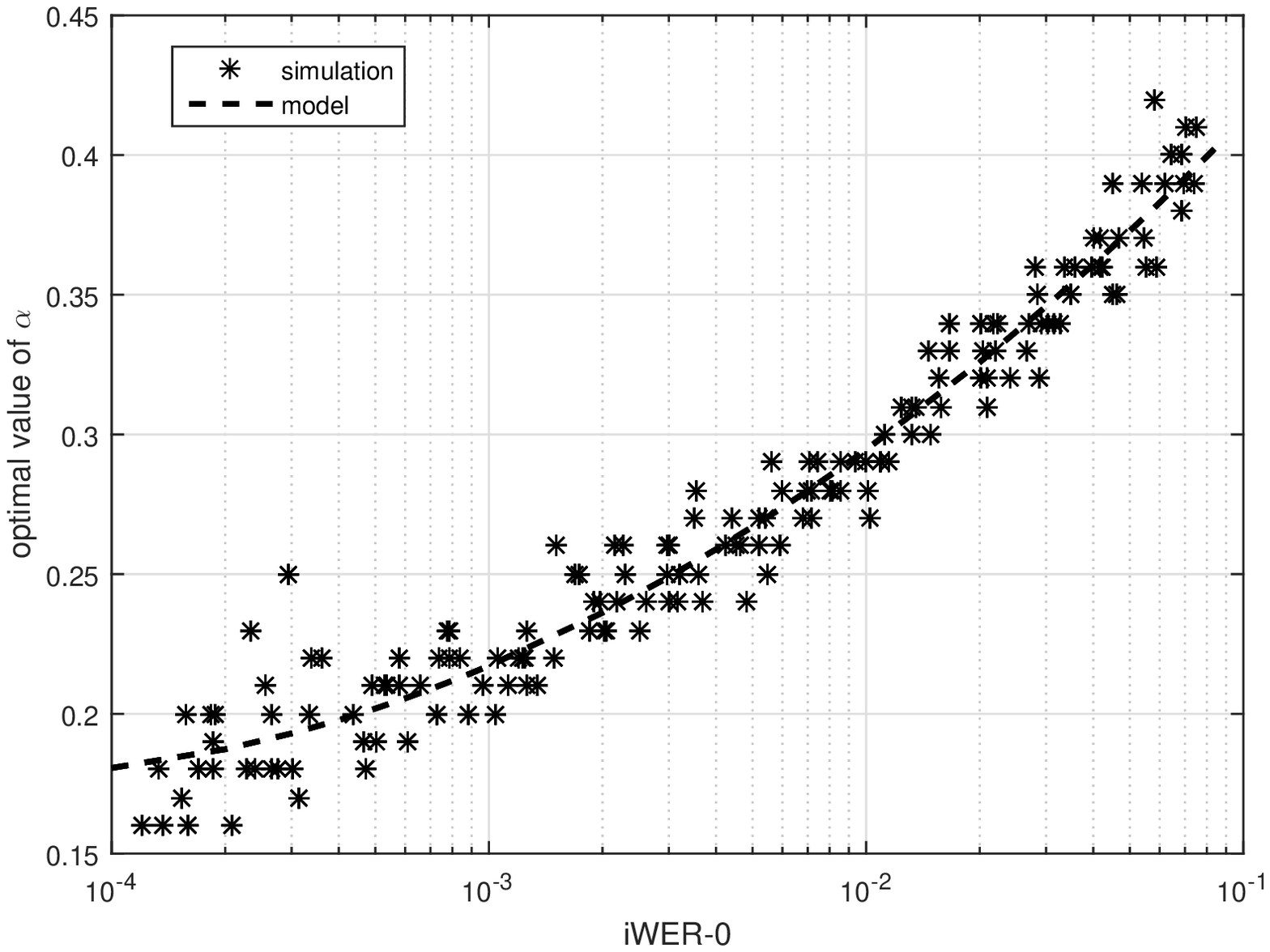}\label{fig:model_alpha}}%
\hfill\subfigure[$\mathbb{E}\left(\text{rk}_{\alpha_{\text{opt}}}(\mathcal{E}_{\mathbf{Y}})\right)$ as function of $\mathtt{i}$WER-$0$]{\includegraphics[width=0.48\linewidth]{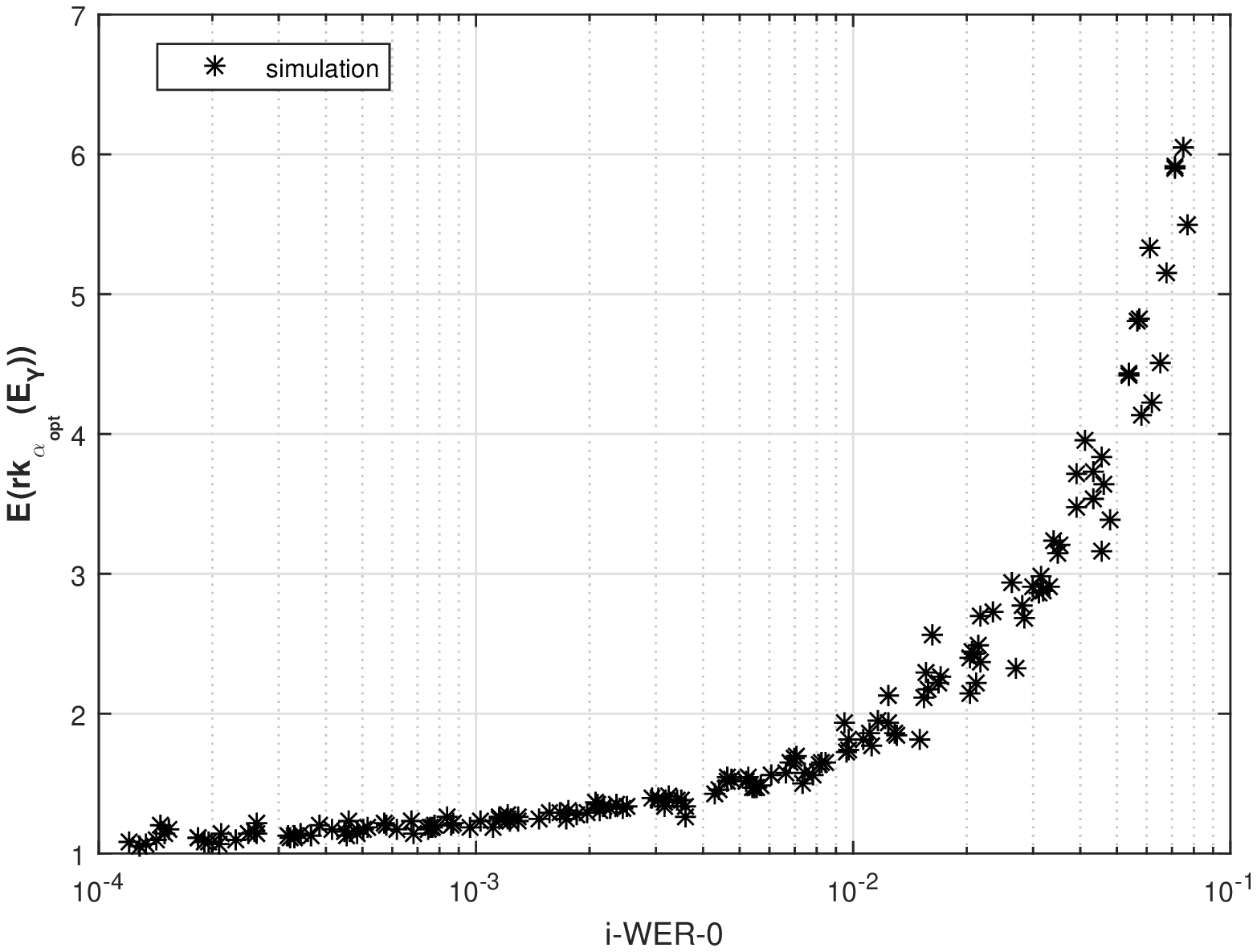}\label{fig:Erk_iwer0}}
\vspace*{5mm}    
    \caption{Optimal values as function of $\mathtt{i}\text{WER}$-$0$ for various code-lengths $N$, coding-rates $R$, and SNR values. Each point corresponds to a different triplet $(N, R, \text{SNR})$.}
\end{figure}

We have determined the $\alpha_{\text{opt}}$ value by Monte-Carlo simulation, for various code parameters $(N, K+r)$ and SNR values. For each pair $(N, K+r)$ and SNR value, we also determined the corresponding $\mathtt{i}$WER-$0$ value, {\em i.e.}, the WER of the SC decoder for a polar code with parameters $(N, K+r)$, as explained in Section~\ref{subsec:wer_lower_bounds}. Precisely, we have considered parameters $(N,K+r)=(256,\{96, 128, 160\}), (512,\{192, 288, 256, 320\}), (1024,\{384, 512, 640\})$, while the SNR values have been chosen such that  $\mathtt{i}\text{WER-}0$ varies from $10^{-4}$ to $10^{-1}$.
Fig.~\ref{fig:model_alpha} shows the scatter plot of  $\alpha_{\text{opt}}$ as a function of $\mathtt{i}$WER-$0$, while Fig.~\ref{fig:Erk_iwer0} shows the scatter plot of $\mathbb{E}\left(\text{rk}_{\alpha_{\text{opt}}}(\mathcal{E}_{\mathbf{Y}})\right)$  as a function of $\mathtt{i}$WER-$0$. 

%

Fig.~\ref{fig:model_alpha} clearly indicates a correlation between $\alpha_{\text{opt}}$ and $\mathtt{i}$WER-$0$ values. Hence, we propose to approximate the $\alpha_{\text{opt}}$ value, by using a quadratic model (in semilog scale):
\begin{equation} \label{eq:model_alpha}
\alpha_{\text{model}}(\mathtt{i}\text{WER-}0)=a_1 \cdot \log(\mathtt{i}\text{WER-}0)^2+a_2 \cdot \log(\mathtt{i}\text{WER-}0) +a_3,
\end{equation}
with the following coefficients providing the best fit to the simulation data:
$$a_1=0.0038,\quad a_2=0.0779,\quad a_3=0.5716$$ 
It is worth noticing that the band of the $\mathbb{E}\left(\text{rk}_{\alpha_{\text{opt}}}(\mathcal{E}_{\mathbf{Y}})\right)$ scatter plot in Fig.~\ref{fig:Erk_iwer0} is very narrow, therefore approximating $\alpha_{\text{opt}}$ by $\alpha_{\text{model}}$ should result in a negligible difference in terms of rank expectation. Fig.~\ref{fig:Erk_iwer0} also demonstrates the effectiveness of the proposed metric in ranking in top positions the bit-flips $\mathcal{E}_{\mathbf{Y}}$.

Finally, we describe below an efficient algorithm to compute $\text{rk}_\alpha(\mathcal{E}_{\mathbf{Y}})$. For this, one needs to determine the number of bit-flips ${\cal E}$, such that $M_\alpha({\cal E}) \geq M_\alpha(\mathcal{E}_{\mathbf{Y}})$. 

Let ${\cal L}_{m} = \{ {\cal E} \mid M_\alpha({\cal E}) \geq m\}$, where $m\in[0,1]$. To determine ${\cal L}_{m}$, we proceed as follows:
\begin{itemize}
\item First, we evaluate $M_\alpha({\cal E}_1)$ for all the bit-flips ${\cal E}_1$ of order $1$, and add to ${\cal L}_{m}$ those bit-flips such that $M_\alpha({\cal E}_1)\geq m$.
\item For $\omega > 1$, we evaluate $M_\alpha({\cal E}_\omega)$ for all the bit-flips ${\cal E}_\omega$ of order $\omega$, such that ${\cal E}_{\omega-1}\in {\cal L}_m$, where ${\cal E}_{\omega-1}$ denotes the bit-flip determined by the first $\omega-1$ elements of ${\cal E}_\omega$. We add to ${\cal L}_{m}$ the bit-flips of order $\omega$, such that $M_\alpha({\cal E}_\omega)\geq m$.
\item The algorithm stops if, at the previous step, no bit-flip of order $\omega$ is added to ${\cal L}_{m}$.
\end{itemize}
Clearly, $\text{rk}_\alpha(\mathcal{E}_{\mathbf{Y}})$ is equal to the number of bit-flips in ${\cal L}_{m}$, for $m = M_\alpha(\mathcal{E}_{\mathbf{Y}})$.

\section{Dynamic SCFlip Decoder}\label{sec:dscflip}

\subsection{D-SCFlip and D-SCFlip-$\omega$ Decoders}\label{subsec:dscflip_saturate}


In this section we introduce a generalized SCFlip decoding algorithm, characterized in that the bit-flip list ${\cal L}_{\text{flip}}$ contains the $T$ bit-flips ${\cal E}$ with highest $M_\alpha({\cal E})$ values, according to the current noise realization $\mathbf{Y}$. In order to avoid the evaluation of $M_\alpha({\cal E})$ for all possible bit-flips ${\cal E}$, the proposed algorithm builds the list ${\cal L}_{\text{flip}}$ on-the-fly, concurrently with the initial SC decoding attempt, and then with each new decoding attempt SC$({\cal E})$. 

The proposed algorithm, referred to as {\em Dynamic SC-Flip} (D-SCFlip), is described in Algorithm~\ref{DSCF}.
The description is similar to the one in Algorithm~\ref{SCF1}, except of the functions $\mathrm{Init}()$ and $\mathrm{Update}()$, 
used to initialize and update the list of bit-flips $\mathcal{L}_{\text{flip}}$ and the list of corresponding metric values $\mathcal{M}_{\text{flip}} \stackrel{\text{def}}{=} \{ M_\alpha({\cal E}) \mid {\cal E} \in \mathcal{L}_{\text{flip}}\}$.

$\mathrm{Init}(\mathcal{L}_{\text{flip}}, \mathcal{M}_{\text{flip}}, \{\text{L}_i\}_{i \in \mathcal{I}})$: this function evaluates $M_\alpha({\cal E})$ for all the bit-flips of order $1$, ${\cal E}=\{i\}$, $i\in{\cal I}$, and orders them according to decreasing value of $M_\alpha({\cal E})$. $\mathcal{L}_{\text{flip}}$ is initialized with the $T$ bit-flips of order $1$ with highest metric values, and the ordered metric values are stored in $\mathcal{M}_{\text{flip}}$. Note that the $M_\alpha({\cal E})$ values computed at this step make use of the LLR values $\{\text{L}_i\}_{i \in \mathcal{I}}$ computed during the initial SC decoding attempt (see Eq.~(\ref{eq:metric_order1})).


\begin{algorithm}[!t]
\caption{D-SCFlip decoder}
\label{DSCF}
\begin{algorithmic}[1]
\Procedure{D-SCFlip}{$\mathbf{Y},\mathcal{I},T$}
	\State $\left(\hat{u}_1^{N}, \{\text{L}_i\}_{i \in \mathcal{I}}\right) \leftarrow $SC$(\emptyset)$ 
	\IIf{CRC($\hat{u}_1^{N}$) = success} $\mathrm{return}\ \hat{u}_1^{N}$; 
	\State {\bf else} 
	$\mathrm{Init}(\mathcal{L}_{\text{flip}}, \mathcal{M}_{\text{flip}}, \{\text{L}_i\}_{i \in \mathcal{I}})$; \EndIIf
	\For {$t=1,\dots,T$}	
		\State $\left(\hat{u}_1^{N}, \{\text{L}[\mathcal{E}_t]_i\}_{i \in \mathcal{I}}\right) \leftarrow $SC$(\mathcal{E}_t)$ 
		\IIf{CRC($\hat{u}_1^{N}$) = success} $\mathrm{return}\ \hat{u}_1^{N}$; 
		\State {\bf else}  $\mathrm{Update}(\mathcal{L}_{\text{flip}}, \mathcal{M}_{\text{flip}}, \{\text{L}[\mathcal{E}_t]_i\}_{i \in \mathcal{I}}, {\cal E}_t)$; \EndIIf 
	 \EndFor
	\State $\mathrm{return}\ \hat{u}_1^{N}$;
\EndProcedure
\end{algorithmic}
\end{algorithm}

\begin{algorithm}[t!]
\caption{Bit-flips update}
\label{updateDSCF}
\begin{algorithmic}[1]
\Procedure{Update}{$\mathcal{L}_{\text{flip}}, \mathcal{M}_{\text{flip}}, \{\text{L}[\mathcal{E}_t]_i\}_{i \in \mathcal{I}}, {\cal E}_t$}
	\For {$i = \mathrm{last}(\mathcal{E}_t)+1,\dots,N$ and $i \in \mathcal{I}$}
		\State ${\cal E} = \mathcal{E}_t\cup \{i\}$; $m = M_\alpha({\cal E})$;
		\If{$m >\mathcal{M}_{\text{flip}}(T)$}
        \State $\mathrm{Insert\_flip}(\mathcal{L}_{\text{flip}}, \mathcal{M}_{\text{flip}}, {\cal E}, m)$;
		\EndIf
	\EndFor
	\State $\mathrm{return}\ (\mathcal{L}_{\text{flip}}, \mathcal{M}_{\text{flip}})$;
\EndProcedure
\end{algorithmic}
\end{algorithm}

$\mathrm{Update}(\mathcal{L}_{\text{flip}}, \mathcal{M}_{\text{flip}}, \{\text{L}[\mathcal{E}_t]_i\}_{i \in \mathcal{I}}, {\cal E}_t)$: After each unsuccessful decoding attempt $\mathcal{E}_t$, $\mathcal{L}_{\text{flip}}$ and $\mathcal{M}_{\text{flip}}$ are updated, as described in Algorithm~\ref{updateDSCF}. Let $\mathcal{E}_t=\{i_1,\dots,i_{\omega_t}\}$, where $\omega_t$ is the order of $\mathcal{E}_t$. The function evaluates $M_\alpha({\cal E})$, for all the all the bit-flips ${\cal E} = \mathcal{E}_t \cup \{i\}$, where $i\in{\cal I}$ and $i > i_{\omega_t}$. In case that $M_\alpha({\cal E}) > \mathcal{M}_{\text{flip}}(T)$, $\mathcal{L}_{\text{flip}}$ and  $\mathcal{M}_{\text{flip}}$ are updated by inserting ${\cal E}$ and $M_\alpha({\cal E})$ into appropriate positions. Since $M_\alpha({\cal E}_t) > M_\alpha({\cal E})$, ${\cal E}$ is necessarily inserted into a position $t'> t$. Note that the $M_\alpha({\cal E})$ values computed at this step make use of the LLR values $\{\text{L}[\mathcal{E}_t]_i\}_{i \in \mathcal{I}}$ computed during the SC(${\cal E}_t$) decoding attempt (see Eq.~(\ref{eq:equation3})). We also note that the initialization of $\mathcal{L}_{\text{flip}}$ and $\mathcal{M}_{\text{flip}}$ can also result from the update procedure of Algorithm~\ref{updateDSCF}, by taking ${\cal E}_t = \emptyset$.

\medskip We now substantiate the ability of the D-SCFlip decoder to explore the bit-flips with highest metric values.
We denote by ${\cal L}_{\mathbf{Y}}=\left\{ {\cal E}_1, \dots, {\cal E}_{T_\mathbf{Y}} \right\} \subset \mathcal{L}_{\text{flip}}$ the ordered list of bit-flips corresponding to the decoding attempts  performed by the D-SCFlip decoder for the current noise realization $\mathbf{Y}$ (not including the initial SC decoding attempt). Hence, $T_\mathbf{Y} \leq T$, since the D-SCFlip decoder stops as soon as a decodig attempt satisfies the CRC. Put differently, ${\cal L}_{\mathbf{Y}}$ is determined by the first $T_\mathbf{Y}$ bit-flips in $\mathcal{L}_{\text{flip}}$, at the moment when the D-SCFlip decoder stops. 

\begin{proposition}\label{prop:dscflip_list}
${\cal L}_{\mathbf{Y}}$ contains the $T_\mathbf{Y}$ bit-flips with the highest $M_\alpha({\cal E})$ values among all the possible bit-flips ${\cal E}$.
\end{proposition}

{\em Proof}. We have to prove that for any bit-flip ${\cal E} = \{i_1,\dots, i_\omega\}$ of order $\omega$, such that $M_\alpha({\cal E}) > M_\alpha({\cal E}_{T_\mathbf{Y}})$, then ${\cal E}\in {\cal L}_{\mathbf{Y}}$. We proceed by induction on $\omega$. For $\omega=1$, the assertion follows from the fact that $\mathcal{L}_{\text{flip}}$ is initialized with the $T$ bit-flips of order $1$ with the highest metric values. For $\omega > 1$, let ${\cal E}' = \{i_1,\dots, i_{\omega-1}\}$. Since $M_\alpha({\cal E}') > M_\alpha({\cal E}) > M_\alpha({\cal E}_{T_\mathbf{Y}})$, it follows from the induction hypothesis that ${\cal E}' \in {\cal L}_{\mathbf{Y}}$. Hence, the decoding attempt SC$({\cal E}')$ is necessarily performed before SC$({\cal E}_{T_\mathbf{Y}})$, and $\mathcal{L}_{\text{flip}}$ is updated after SC$({\cal E}')$ by evaluating the bit-flips of order $\omega$ that contains ${\cal E}'$. During this update, ${\cal E}$ is  added to $\mathcal{L}_{\text{flip}}$, in a position that necessarily precedes that of ${\cal E}_{T_\mathbf{Y}}$. Therefore, ${\cal E}\in {\cal L}_{\mathbf{Y}}$, which completes the proof. \hfill$\square$

\medskip Finally, we denote by D-SCFlip-$\omega$ the decoder obtained by restricting $\mathcal{L}_{\text{flip}}$ to  bit-flips of order less than or equal to $\omega$. It has a similar description to the one provided in Algorithm~\ref{DSCF}, but the update procedure  $\mathrm{Update}(\mathcal{L}_{\text{flip}}, \mathcal{M}_{\text{flip}}, \{\text{L}[\mathcal{E}_t]_i\}_{i \in \mathcal{I}}, {\cal E}_t)$ is only executed if the order of ${\cal E}_t$ is less than $\omega$ (thus, no bit-flip of order greater than $\omega$ is added to $\mathcal{L}_{\text{flip}}$). Similarly to Proposition~\ref{prop:dscflip_list}, it can be seen that the D-SCFlip-$\omega$ decoder explores the bit-flips ${\cal E}$ of order less than or equal to $\omega$,  with the $T_\mathbf{Y}$ highest $M_\alpha({\cal E})$ values. The purpose of the D-SCFlip-$\omega$ decoder is to assess the effectiveness of the proposed metric in approaching the performance of the ideal $\mathtt{i}$SCFlip-$\omega$ decoder, defined in Section~\ref{subsec:wer_lower_bounds}. We also note that limiting the maximum bit-flip order may have some practical payoffs, or be imposed by some practical constraints ({\em e.g.}, related to hardware implementation), but such considerations are beyond the scope of this paper. 

\subsection{Practical implementation}
\label{subsec:practical_implem}

This section discusses two practical simplifications, which allow reducing the computational cost of implementing the proposed D-SCFlip decoder. First, to reduce the computational cost associated with new decoding attempts, the following proposition determines the position from which the SC decoding need to be restarted.  

\begin{proposition}
Let $\mathcal{E}_{1}=\{i_1, i_2 \dots, i_{\omega_{1}} \}$ and $\mathcal{E}_{2}=\{j_1, j_2 \dots, j_{\omega_{2}} \}$ be two bit-flips, and $1\leq \omega \leq\min(\omega_{1}, \omega_{2})$ be such that $i_{\omega'} = j_{\omega'}$ for any $\omega' < \omega$, and $i_{\omega} \neq j_{\omega}$. Let $k=\min(i_\omega, j_\omega)$. Then  SC($\mathcal{E}_{1}$) and SC($\mathcal{E}_{2}$) decoding attempts are strictly identical in terms of LLRs and hard-decision estimates until index $k$, where they differ only by the hard-decision estimate of the bit $u_k$.
\end{proposition}

As a consequence, assuming that SC($\mathcal{E}_{1}$) and SC($\mathcal{E}_{2}$) are two successive decoding attempts,  the latter may start from the index $k+1$,  after the hard-decision estimate of $u_k$ has been flipped. 


%
%

The following proposition, which follows  from Eq.~(\ref{eq:equation3}), allows reducing the computational complexity of the $\mathrm{Update}$ procedure. It allows avoiding the computation of the metric values $m=M_{\alpha}({\cal E})$ (see Algorithm~\ref{updateDSCF}), for bit-flips ${\cal E} = {\cal E}_t \cup \{i\}$ which would not be inserted in the list anyway. 

\begin{proposition}
Consider the update procedure $\mathrm{Update}(\mathcal{L}_{\text{flip}}, \mathcal{M}_{\text{flip}}, \{\text{L}[\mathcal{E}_t]_i\}_{i \in \mathcal{I}}, {\cal E}_t)$, after some decoding attempt SC$({\cal E}_t)$, with $\mathcal{E}_t=\{i_1,\dots,i_{\omega_t}\}$. 
For any $i > i_{\omega_t}$, let 
\begin{equation}
\Pi({\cal E}_t, i) = \prod_{\underset{j \in \mathcal{I}}{j=i_{\omega_t}+1}}^{i-1} \left( \frac{1}{1+\exp{(-\alpha|\text{L}[{\cal E}_t]_j|)}} \right)
\end{equation}
Then:
\begin{itemize}
\item[(i)] $M_\alpha(\mathcal{E}_t \cup \{i\}) = \displaystyle M_\alpha(\mathcal{E}_t)\cdot \Pi({\cal E}_t, i) \cdot \frac{1}{1+\exp{(\alpha |\text{L}[\mathcal{E}_{t}]_{i})|)}}$\vspace*{2mm}
\item[(ii)] 
%
%
Let $i_{\omega_t} < k \leq N-1$  be the last (highest) value such that $M_{\alpha}({\cal E}_t) \cdot \Pi({\cal E}_t, k) \geq \mathcal{M}_{\text{flip}}(T)$ (note that for $k=i_{\omega_t}+1$, $M_{\alpha}({\cal E}_t) \cdot \Pi({\cal E}_t, k) = M_{\alpha}({\cal E}_t) \geq \mathcal{M}_{\text{flip}}(T)$). Then, $\mathcal{M}_{\text{flip}}(T) > M_{\alpha}({\cal E}_t) \cdot \Pi({\cal E}_t, i) > M_\alpha(\mathcal{E}_t \cup \{i\})$, for any $i= k+1,\dots,N-1$. In particular, the {\bf for} loop in Algorithm~\ref{updateDSCF} (line 2) can be restricted to values $i=i_{\omega_t}+1,\dots,k$.   
\end{itemize}
\end{proposition}

\subsection{Numerical Results}\label{subsec:numerical_results}

All the simulation results presented in this section assume a BI-AWGN channel. Concatenated CRC-polar codes use $(r=16)$-bits CRC, with generator polynomial $g(x)=x^{16}+x^{15}+x^{2}+1$. The set $\mathcal{I}$ is optimized for each SNR value, according to the Gaussian Approximation method presented in \cite{trifonov2012efficient}.

\begin{figure}[t!]
    \centering
    \includegraphics[width=0.6\linewidth]{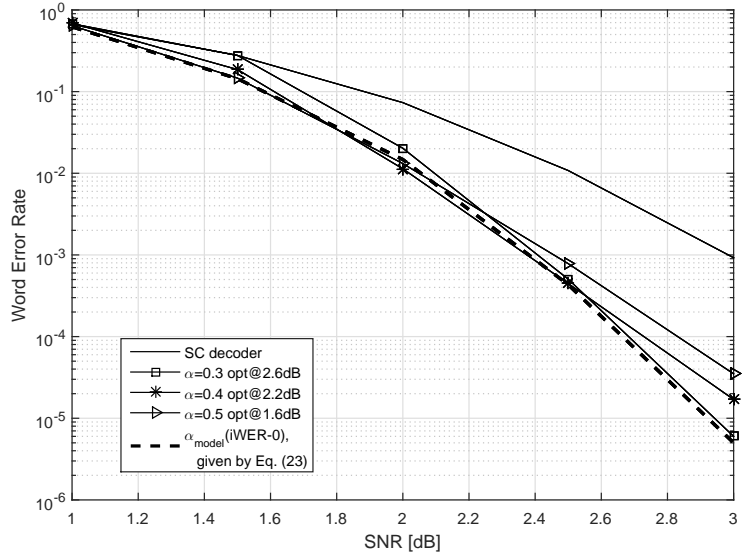}
    \caption{Impact of $\alpha$ on the performance of D-SCFlip for $T=20$ for a code $(N,K+r)=(1024,512+16)$}
    \label{fig:wer_impact_alpha}
\end{figure}

We start by investigating the impact of the parameter $\alpha$ on the decoding performance of the D-SCFlip decoder. Fig.~\ref{fig:wer_impact_alpha} shows the WER performance of the D-SCFlip decoder for a concatenated CRC-Polar code with parameters $(N,K+r)=(1024,512+16)$, and several {\em fixed} $\alpha$ parameters, where {\em fixed} means that the same $\alpha$ parameter is used for all the SNR values. Each $\alpha$ parameter corresponds to the optimal $\alpha_{\text{opt}}$ value for a particular SNR (Section~\ref{subsec:optimal_alpha}), which is indicated in the legend.  The maximum number of extra decoding attempts ({\em i.e.}, not including the initial SC decoding attempt) is set to $T = 20$. For comparison purposes, the WER performance of the SC decoder for the $(N,K)=(1024,512)$ polar code is also shown. The dashed curve plots the WER performance using the $\alpha_{\text{model}}(\mathtt{i}\text{WER-}0)$ value, which varies with the SNR. Precisely, for each SNR value we first determine {\em offline} the corresponding $\mathtt{i}$WER-$0$ value, by using the OA-SC decoder, then the value of $\alpha_{\text{model}}(\mathtt{i}\text{WER-}0)$, according to Eq.~(\ref{eq:model_alpha}). 
The figure highlights the performance loss -- in the low, medium or high SNR regime -- when a fixed $\alpha$ value is used throughout the whole range of SNR values. It also demonstrates the effectiveness of the proposed model, since the $\alpha_{\text{model}}$ curve matches the envelope of the curves with a fixed $\alpha$.
For all the simulation results presented in the remaining of this section, we shall assume that 
$\alpha = \alpha_{\text{model}}(\mathtt{i}\text{WER-}0)$.

\begin{figure}[t!]
    \centering
    \includegraphics[width=0.6\linewidth]{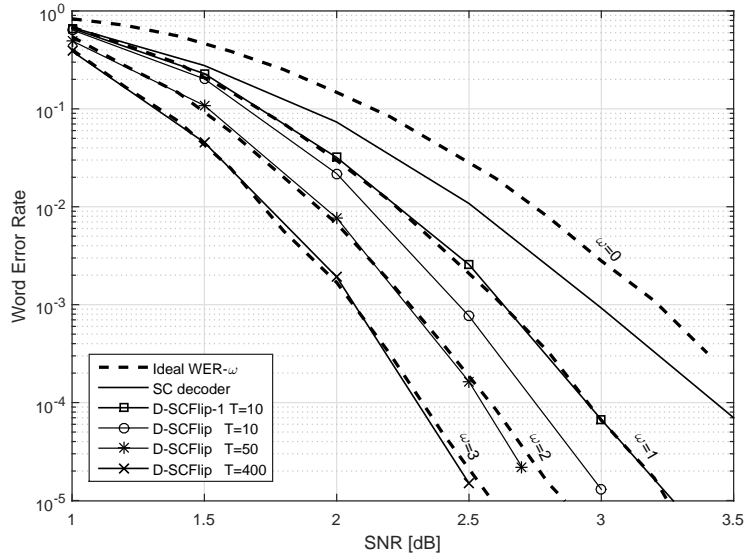}
    \caption{Performance of D-SCFlip decoder for several values of $T$ and a code $(N,K+r)=(1024,512+16)$}
    \label{fig:wer_num_attempts}
\end{figure}

Fig.~\ref{fig:wer_num_attempts} shows the WER performance of the D-SCFlip for $T \in \{10,50,400\}$, for the concatenated CRC-Polar code with parameters $(N,K+r)=(1024,512+16)$. The values of $T$ have been chosen such that the D-SCFlip performance is close to or outperforms the ideal performance $\mathtt{i}$WER-$\omega$ for $\omega=1,2$ and $3$ respectively, thus proving the ability of the proposed both metric and decoder to correct higher-order noise realizations. The impact of saturating to a low value of $\omega$ is also shown by considering a D-SCFlip-($\omega=1$) decoder with $T=10$, for which the performance tightly approaches the ideal performance $\mathtt{i}$WER-1.

\begin{figure}[t!]
    \centering
    \includegraphics[width=0.6\linewidth]{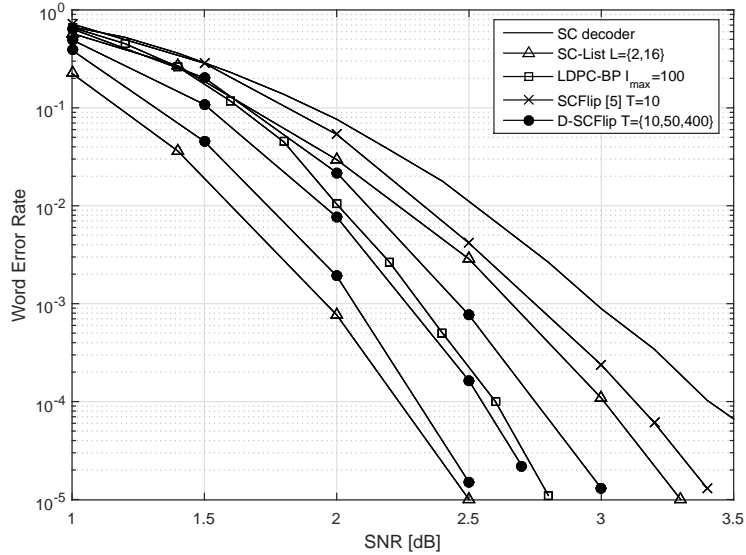}
    \caption{Comparison between D-SCFlip decoder and BP decoder of LDPC codes and SC-List decoders of polar codes at length $N=1024$ and rate $R=0.5$}
    \label{fig:ComparDec}
\end{figure}

Fig.~\ref{fig:ComparDec} provides a comparison of the D-SCFlip, SCFlip \cite{afisiadis2014low}, and SCL decoders for concatenated CRC-polar codes, with $(N, K+r) = (1024, 512+16)$. The D-SCFlip decoder has a maximum number of extra decoding attempts $T=\{10,50,400\}$. The performance of the state-of-the-art SCFlip \cite{afisiadis2014low} is given for $T=10$. However, as this decoder is actually a D-SCFlip-1 with $\alpha=+\infty$ (see Section~\ref{subsec:impact_alpha}), even for higher values of $T$, its performance is still lower bounded by the ideal performance $\mathtt{i}$WER-1. On the contrary, the D-SCFlip decoder performance significantly improves with increasing $T$ values,  outperforming the SCFlip decoder by $0.4$\,dB for $T=10$, and $0.8$\,dB for $T=400$ (at $\text{WER} = 10^{-4}$). 

For $T=400$, the proposed D-SCFlip decoder closely approaches the performance of the SCL decoder with $L=16$. Even though the maximum number of additional attempts ($T=400$) used by the D-SCFlip decoder is considerably higher than the size of the list ($L=16$) used by the SCL decoder, it should be understood that the trade-off of the D-SCFlip decoder is different: it offers a low computational complexity, especially in moderate to high SNR regime, since additional decoding attempts are performed only in case the SC decoding fails.

For comparison purposes, we have also included in Fig.~\ref{fig:ComparDec} the WER performance of a $(3,6)$-regular LDPC code, with $(N,K) = (1024, 512)$, under Belief Propagation (BP) decoding. The  LDPC code is constructed by using the Progressive Edge Growth (PEG) algorithm \cite{PEG_LDPC}, and has girth $g=8$. The  maximum number of iterations for the BP decoding is set to $100$, since for higher values the performance improvement is actually  negligible. It can be observed that the BP decoder is outperformed by the D-SCFlip decoder for $T\geq 50$. 

\begin{figure}[t!]
    \centering
    \includegraphics[width=0.6\linewidth]{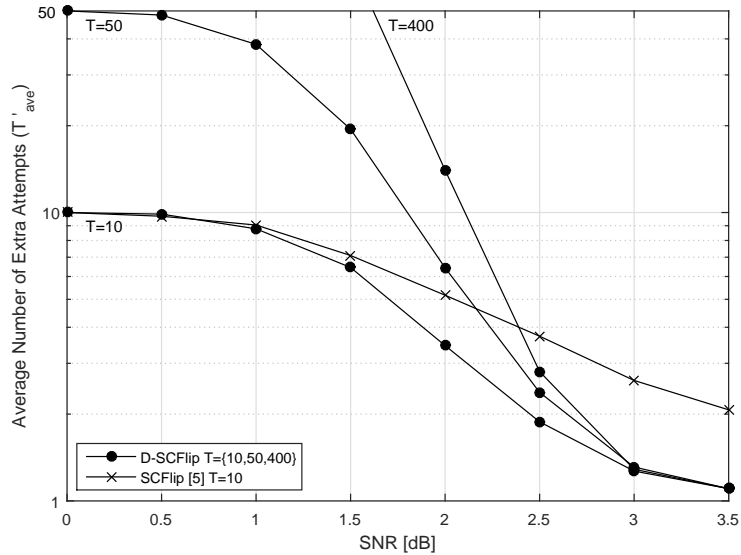}
    \caption{Average number of extra attempts ($T'_{\text{ave}}$) for D-SCFlip decoders and $N=1024$ and rate $R=0.5$}
    \label{fig:Average_Complexity}
\end{figure}

The average number of extra decoding attempts performed by the D-SCFlip in case the SC decoder fails, denoted by $T'_{\text{ave}}$, is shown in Fig.~\ref{fig:Average_Complexity}. One can observe that $T'_{\text{ave}}$ quickly drops and approaches $1$ for high SNR values, demonstrating the effectiveness of the proposed D-SCFlip decoder, and implicitly of the proposed bit-flip metric, in finding the higher-order bit-flips that correct the actual noise realization. Comparing with the SCFlip from \cite{afisiadis2014low}, the proposed D-SCFlip decoder requires a smaller number of extra decoding attempts at high SNR, thus resulting in a lower computational complexity, while providing a significant gain in terms of WER performance.

Finally, let $T_{\text{ave}}$ denote the {\em overall average} number of extra decoding attempts, {\em i.e.}, averaged over all the cases, irrespective of the SC decoder status (successful or not). It follows that $T_{\text{ave}} = T'_{\text{ave}} \text{WER}_{\text{SC}}$, where $\text{WER}_{\text{SC}}$ denotes the WER of the SC decoder. 
Compared to the SC decoding, D-SCFlip decoding results in an increase of both the average computational complexity\footnote{We do not take into account the practical simplifications proposed in Section~\ref{subsec:practical_implem}, and assume that the computational complexity of each new decoding attempt is the same as the one of the initial SC decoding. The computational complexity of the $\mathrm{Update}$ procedure is not taken into account, since it is linear in $N$, and thus negligible with respect to the computational complexity of SC.} and average decoding latency by a factor of only $(1+T_{\text{ave}})$, where the term $1$ in the parenthesis accounts for the initial SC decoding attempt. Note also that the contribution of $T_{\text{ave}}$  actually becomes negligible in the waterfall region of the SC decoder.  As a matter of comparison, the computational complexity of the SCL decoder, with a list of size $L$, is $L$ times higher than the one of the SC decoding, while they both have the same decoding latency. D-SCFlip considerably reduces the computational complexity, by relying on  successive -- rather than parallel -- decoding attempts, coupled with a judicious choice of the latter ones. While this results in a variable decoding latency, with worst case latency given by the maximum number of decoding attempts $T$, the average decoding latency is nearly the same as the latency of the SC decoding.

\section{Conclusion}

In this paper, we investigated a Generalized SCFlip decoding for polar codes, 
characterized by $T$ new decoding attempts, where one or several positions are flipped from the standard SC decoding.
First, we studied the WER performance of an ideal Generalized SCFlip decoder, with maximum bit-flip order $\omega$, which revealed potential for significant improvements, enabled by the use of higher-order bit-flips.  Subsequently, we concentrated on proposing a practical method to take advantage of the benefits offered by the use of higher-order bit-flips, which led to two complementary improvements. First, a new metric was proposed, suited to  bit-flips of any order, and optimized such that the sequential aspect of successive cancellation decoder is accurately taken into consideration. We also provided an analysis of the impact of the parameter $\alpha$ used within the proposed metric, and proposed an empirical model to estimate its optimal value as a function of $\mathtt{i}$WER-$0$, which can be easily evaluated by using the density evolution technique. Secondly, we investigated a method to  dynamically build the bit-flips list $\mathcal{L}_{\text{flip}}$, so that to guarantee that new decoding attempts are performed by decreasing probability of success, according to the proposed metric. The resulting D-SCFlip algorithm was shown to offer a substantial gain in terms of WER performance, as compared to the state-of-the-art SCFlip decoder, while having a lower computational complexity. Finally, we showed that the D-SCFlip decoder is an interesting variable-latency approach, which provides a different trade-off compared to SCL decoding of polar codes, by keeping the computational complexity close to the one of the SC decoder, while providing decoding performance close to SCL decoding with list size $L=16$.

\bibliographystyle{IEEEtran}
\bibliography{biblio_journal}

\end{document}